\documentclass{aastex}

\begin{document}

\title{A 10 $\mu$m Search for Inner-Truncated Disks Among Pre-Main-Sequence 
Stars with Photometric Rotation Periods} 

\author{Keivan G. Stassun\altaffilmark{1} and Robert D. Mathieu\altaffilmark{1}}
\author{Frederick J. Vrba\altaffilmark{2}}
\author{Tsevi Mazeh\altaffilmark{3}}
\author{Arne Henden\altaffilmark{2}}

\altaffiltext{1}{Astronomy Dept., University of Wisconsin, 475 N.
Charter St., Madison, WI 53706; keivan@astro.wisc.edu}
\altaffiltext{2}{US Naval Observatory, Flagstaff Station, Flagstaff, AZ 86002-1149}
\altaffiltext{3}{Wise Observatory, Tel Aviv University, Tel Aviv, Israel}

\shorttitle{Truncated Disks Among TTS}
\shortauthors{Stassun et al.}

\begin{abstract}
We use mid-IR (primarily 10 $\mu$m) photometry as a diagnostic for the
presence of disks with inner cavities among 32 pre-main sequence stars in Orion
and Taurus-Auriga for which rotation periods are known and which do
not show evidence for inner disks at near-IR wavelengths. 
Disks with inner cavities are
predicted by magnetic disk-locking models that seek to explain the 
regulation of angular momentum in T Tauri stars.
Only three stars in our sample show evidence for excess mid-IR emission.
While these three stars may possess
truncated disks consistent with magnetic disk-locking
models, the remaining 29 stars in our sample do not.
Apparently, stars lacking near-IR excesses in general do not 
possess truncated disks to which they are magnetically coupled.

We discuss the implications of this result for the
hypothesis of disk-regulated angular momentum. 
Evidently, young stars can exist as slow rotators without the aid of 
present disk-locking, and there exist very young stars already rotating
near breakup velocity whose subsequent angular momentum
evolution will not be regulated by disks. 
Moreover, we question whether disks, when present, truncate in the 
manner required by disk-locking scenarios. 
Finally, we discuss the need for rotational evolution models 
to take full account of the 
large dispersion of rotation rates present at 1 Myr, which
may allow the models to explain the rotational evolution of low-mass
pre-main sequence stars in a way that does not depend upon
braking by disks.

\end{abstract}


\section{Introduction}
A keystone of the current paradigm of rotational evolution among low-mass
pre-main-sequence (PMS) stars is the regulation of stellar angular
momentum by circumstellar disks.
Indeed, most rotational evolution models today rely chiefly upon some 
form of disk-regulated stellar angular momentum to connect
the rotational properties of T Tauri stars (TTS) to those of zero-age
main sequence (ZAMS) stars \citep{bouvier97,krishnamurthi,cameron95}.
The theoretical edifice supporting the disk-regulation paradigm has been
developed by several researchers 
\citep{konigl91,ostrikershu,yi95,cameron93,ghosh},
wherein a steady-state, magnetically mediated, star-disk coupling 
mechanism keeps the star at 
roughly constant angular velocity for the lifetime of the disk,
preventing the star from spinning up as it contracts.
This magnetic disk-locking model provides 
an attractive solution to 
the long-standing puzzle of how stars deplete angular momentum
as they evolve to the main sequence \citep{stauffhart}. 

Observational investigations of disk-regulated angular momentum have
yielded mixed results. In support of disk-locking, several authors have, 
either directly or indirectly, linked disks to slow stellar rotation. 
For example, Herbst and collaborators \citep[and references therein]{ch}
have reported the distribution of TTS rotation periods in the Orion
Nebula Cluster (ONC) to be distinctly bimodal. The slow rotators
($P_{\rm rot} > 4$ days), they suggest, represent those stars that are
presently, or were until recently, disk-locked. More direct support
has been presented by \citet{edwards93}, who reported a correlation
between the presence of excess near-IR emission and slow stellar 
rotation among 34 TTS in Taurus and Orion. Considering stars
in their sample later than K5, those authors found excess near-IR 
emission among none of the rapid rotators but among half of the slow
rotators. \citet{bouvier93} reported similar results among 26 
TTS in Taurus-Auriga. They found that the classical TTS (CTTS; 
those showing evidence of active accretion and circumstellar disks) 
in their sample rotate more slowly on average than their 
weak-lined (WTTS) counterparts. 
In contrast to these studies, we did not find strong evidence
for a bimodal distribution of rotation periods, nor did we find evidence 
for a linkage between disks and slow stellar rotation, in our own study
of disks and stellar rotation in Orion 
\citep[hereafter Paper~I; see also Stassun et al.\ 2000 for a concise 
summary]{stass}.
Among 254 stars with $0.1 < P_{\rm rot} < 8$ days, we found the 
distribution of rotation periods to be statistically consistent with a 
uniform distribution. 
Moreover, we found near-IR excesses to be present with roughly equal 
frequency among both slow and rapid rotators when we combined our 
rotation periods with near-IR photometry from the literature 
\citep{hill98} for more than 100 stars in our sample. 

\citet{herbst99} have shown that subdividing the ONC sample by mass
can reconcile some of these contradictory findings.
They find that the bimodal distribution of rotation periods in Orion 
holds true when they consider only stars with
$M_\star > 0.25 {\rm M}_\odot$, i.e.\ earlier than about M3. 
For these stars, they also find a weak but statistically significant
correlation between
near-IR excess and stellar rotation period; the slower rotators tend
to exhibit larger near-IR excesses. 

Whether these IR observations of TTS in fact agree with
the predictions of disk-locking models is an important question that 
has yet to be fully explored.
While the various prescriptions for magnetic star-disk coupling 
differ in detail, a key prediction of the models is that the
stellar magnetosphere will truncate the disk at an inner radius, $R_{\rm trunc}$,
that is very nearly equal to the co-rotation radius, $R_{\rm c}$
(the distance from the star at which disk material rotates with Keplerian 
angular velocity equal to the stellar angular velocity).
Consequently, stars with $R_{\rm c} \gg R_\star$ (i.e. slow rotators) 
should have $R_{\rm trunc} \gg R_\star$, and so are {\it not} expected to 
show their disks at near-IR wavelengths (e.g.\ \citet{meyer97}). 
Ironically, while previous authors have upheld the disk-locking hypothesis 
by linking slow stellar rotation to the presence of excess near-IR
emission, it is the
slow rotators that {\it lack} near-IR excesses that may be
most consistent with the disk-locking scenario. 
An important question, therefore, 
is whether slow rotators lacking near-IR excesses indeed possess 
truncated disks, or if they are simply diskless. 

A similar question applies with respect to
{\it rapid} stellar rotation as well.  We reported in Paper~I the 
existence of very rapidly rotating ($P_{\rm rot} < 2$ days)
TTS in the ONC, some with rotation periods 
corresponding to breakup velocity. Roughly one-third of the rapid rotators 
in the ONC do 
not evince excess emission in the near-IR, indicating that they do
not possess inner disks.
What is to be the 
rotational fate of these apparently diskless PMS ultra-fast rotators, 
and what implications do they carry for rotational evolution models? 
At $\lesssim 1$ Myr \citep{hill97}, these stars still face contraction in size by 
factors of 5--10, yet {\it already rotate at or near breakup velocity}.
If these stars are truly diskless as the near-IR data suggest, then
they must shed most of their angular momentum
prior to the main sequence without the aid of disk-locking.

In this paper we present new mid-IR (primarily $N$-band)
photometry for a sample of low-mass TTS in the ONC and in Taurus-Auriga
for which rotation periods have been determined and which do not 
show excess emission in the near-IR. 
We supplement these new
data with existing data from the literature to form a sample of 32 stars,
roughly evenly divided between slow ($P_{\rm rot} > 4$ days) and rapid rotators.
Our aim is two-fold: First, 
we wish to determine whether slow rotators lacking near-IR excesses 
possess truncated disks, consistent with disk-locking models, or if 
these stars are simply diskless. Second, we wish to determine whether 
{\it rapid} rotators lacking near-IR excesses are in fact diskless, or
if they possess truncated disks to which they may be coupled in
a non-steady-state and which may influence their subsequent rotational evolution.

The data, and 
the methods we employ in their analysis, are described in \S 2. 
The results of our search for truncated disks are presented in \S 3.
We find evidence for excess mid-IR emission among only three of the stars 
in our sample. These stars may possess disks with 
$R_{\rm trunc} \approx R_{\rm c}$, consistent with disk-locking models. 
However, no disks are evident at mid-IR wavelengths among the remaining 
29 stars in our sample; these stars are evidently diskless.
It thus appears that many stars can exist as slow rotators at 1 Myr without 
the aid of present disk-locking. In addition, there exist very
young stars already rotating near breakup velocity whose subsequent
angular momentum evolution will not be regulated by disks.
In \S 4 we discuss the implications of these results both for
the hypothesis of disk-regulated angular momentum and for rotational
evolution models of low-mass PMS stars. 
We summarize our conclusions in \S 5.

\section{Data and Methods}
Our goal is to search for truncated circumstellar disks among
TTS with known rotation periods but without evidence for disks in
the near-IR. To do this, we wish to search for the presence of 
``excess" flux (emission above photospheric) at mid-IR wavelengths
and to compare
the observed spectral energy distribution (SED) of each star to a
model SED of a star with a truncated disk. 
Thus, we formed the sample
for this study by selecting PMS stars in the ONC and Taurus-Auriga
star-forming regions that satisfy the following criteria: 
(1) known rotation periods, 
(2) near-IR (i.e.\ $JHK$) colors consistent with photospheric values, and
(3) basic stellar data available (e.g.\ effective temperatures, radii, 
luminosities, extinctions) for computing model SEDs. ONC rotation
periods were taken from \citet{ch} and Paper~I, while Taurus-Auriga
rotation periods were taken from \citet{bouvier95} and \citet{grankin}.
We identified stars lacking in near-IR excess emission on the basis of
the near-IR excess-emission diagnostics determined by \citet{hill97} 
and \citet{edwards93} for stars in the ONC, and by \citet{strom89} 
and \citet{edwards93} for stars in Taurus-Auriga. Following the 
criteria discussed in those papers, we selected only sources 
with either $\Delta (I-K) < 0.3$ mag \citep{hill97}, 
$\Delta (H-K) < 0.1$ mag
\citep{edwards93}, or $\Delta K < 0.1$ dex \citep{strom89}.
We consider only stars with derived masses of 
$M_\star > 0.25 {\rm M}_\odot$, as \citet{herbst99} have recently
suggested that this is the stellar mass above which rotational and
infrared signatures of disk-locking are observable at $\sim 1$ Myr.

Here we describe the near- and mid-IR photometry newly obtained by us,
as well as the visible to mid-IR data from the literature, that we 
combine to form observed SEDs. We then briefly describe the method
we employ to generate model SEDs. We compare the observed and model
SEDs in the following section.

\subsection{New IR photometry}
Forty ONC stars met the criteria for selection discussed above, none
of which have mid-IR photometry in the literature.
We selected ONC targets for our mid-IR observations
primarily on the basis of rotation period, attempting to observe
an equal number of rapid and slow rotators. We also gave some
preference to brighter stars, thereby minimizing exposure times 
and maximizing the number of targets we could observe. Twenty-three
Taurus-Auriga stars met the selection criteria, and of these only
eight do not already have mid-IR photometry available in the literature.
We selected Tau-Aur targets for our mid-IR observations entirely
on the basis of rotation period. 

We obtained new $N$-band (10.8 $\mu$m) photometry of 11 TTS in the ONC
and 5 TTS in Taurus-Auriga using the OSCIR instrument\footnote{OSCIR 
is a mid-IR imager and spectrometer, optimized for 8--25
$\mu$m, built at the University of Florida under the direction of
Charles Telesco. See 
\anchor{http://www.astro.ufl.edu/oscir/}{http://www.astro.ufl.edu/oscir/}
for more information about OSCIR.}
on the 4-meter telescope at CTIO over the three nights 1998 Dec 4--6.
The OSCIR imager has a plate scale of 0{\farcs}183 pixel$^{-1}$
and a field-of-view of $23\arcsec \times 23\arcsec$. We used a standard
chop-nod observing mode, with a chop frequency of 5 Hz and throw of
23\arcsec N-S. In addition, for 2 TTS in the ONC (one of which was included 
in the $N$-band observations), we obtained photometry 
in the $L'$ (3.8 $\mu$m) and $M$ (4.8 $\mu$m\footnote{These observations 
were performed with the IRTF narrow-band $M$ filter ($\lambda_0 = 4.77$ 
$\mu$m, $\Delta\lambda = 0.23$ $\mu$m). Narrow-band $M$ magnitudes
are expected to be within a few percent of standard $M$ magnitudes.})
passbands with the NSFCAM imager at the IRTF in service-observing mode 
on 1998 Feb 7. NSFCAM is a 1--5 $\mu$m imager with a $256 \times 256$ 
InSb detector. The instrument was used in its 0{\farcs}15 pixel$^{-1}$ mode.
A log of these mid-IR photometric observations is presented in 
Table~\ref{table1}. 

\objectname[]{$\alpha$ Ori} and \objectname[]{$\alpha$ Tau} were used
as flux standards for the OSCIR $N$-band observations, and 
\objectname[]{HD 40335} and \objectname[]{HR 1552} were used as flux
standards for the NSFCAM $L'M$ observations\footnote{Standard-star fluxes from
\citet{tokunaga}, \citet{elias82}, and from the IRTF catalog of bright 
infrared standard stars.}.
Standard-star measurements were obtained before
and after each target-star observation. After correcting for extinction,
standard-star fluxes obtained with OSCIR show an RMS scatter of 3--5\%
during each of the three nights. Standard-star fluxes obtained with 
NSFCAM display an RMS scatter of 3\%. On the advice of S. Fisher 
(priv. comm.), we conservatively adopt calibration uncertainties of 
10\% in our OSCIR measurements.

\begin{deluxetable}{llccc}
\scriptsize
\tablecolumns{5}
\tablecaption{Log of New Mid-IR Observations\label{table1}}
\tablehead{
\colhead{UT Date} & \colhead{Instrument/Telescope} & \colhead{Star} &
\colhead{Filters} & \colhead{Exp.\ Time On-Source} \\
 & & & & \colhead{(sec)} }
\startdata
1998 Feb 7 & NSFCAM/IRTF & \objectname[]{JW 352}\tablenotemark{2} & $L'$, $M$ & 5, 20 \\*
 & & \objectname[]{JW 567} & $L'$, $M$ & 5, 20 \\
1998 Dec 4 & OSCIR/CTIO 4-m & \objectname[]{JW 157}\tablenotemark{1} & $N$ & 960 \\*
 & & \objectname[]{JW 589}\tablenotemark{1} & $N$ & 720 \\*
 & & \objectname[]{TAP 41}\tablenotemark{1} & $N$ & 450 \\*
 & & \objectname[]{JW 794}\tablenotemark{1} & $N$ & 900 \\*
 & & \objectname[]{JW 641}\tablenotemark{1} & $N$ & 450 \\
1998 Dec 5 & OSCIR/CTIO 4-m & \objectname[]{JW 790} & $N$ & 600 \\*
 & & \objectname[]{V826 Tau}\tablenotemark{1} & $N$ & 360 \\*
 & & \objectname[]{IW Tau}\tablenotemark{1} & $N$ & 600 \\*
 & & \objectname[]{TAP 45}\tablenotemark{1} & $N$ & 1050 \\*
 & & \objectname[]{JW 544}\tablenotemark{1} & $N$ & 300 \\*
 & & \objectname[]{JW 330}\tablenotemark{1} & $N$ & 975 \\
1998 Dec 6 & OSCIR/CTIO 4-m & \objectname[]{JW 866} & $N$ & 600 \\*
 & & \objectname[]{V830 Tau}\tablenotemark{1} & $N$ & 600 \\*
 & & \objectname[]{JW 352}\tablenotemark{2} & $N$ & 600 \\*
 & & \objectname[]{JW 669} & $N$ & 300 \\*
 & & \objectname[]{JW 926} & $N$ & 370 \\
\enddata
\tablenotetext{1}{Contemporaneous $JHK$ photometry also obtained for this target.}
\tablenotetext{2}{Note that \objectname[]{JW 352} appears twice in this table.}
\end{deluxetable}

We also obtained contemporaneous photometry in $JHK$ for 11 of the
17 targets in Table~\ref{table1} so as to minimize the effects of near-IR
variability on our SEDs. Each star was observed on 1998 Dec 3 and again
on 1998 Dec 8 (except \objectname[]{TAP 45}, which was only observed 
on 1998 Dec 8) with the IRCAM instrument on the
61-inch telescope at the US Naval Observatory. 
IRCAM uses a Rockwell $256 \times 256$ HgCdTe array operating at 77K.
At the USNO 61-inch telescope it has a pixel scale of $0{\farcs}54$
with an approximate field of view of $2.'3 \times 2.'3$.
The resultant $JHK$ photometry is on the CIT system \citep{elias82}
based on calibration observations of 15 standard stars each night to
measure atmospheric extinction and system color terms.
Formal uncertainties in these measurements are typically $\le 0.04$ mag in
all three filters, with no uncertainty larger than 0.06 mag, except 
for \objectname[]{JW 589}, for which 
we quote uncertainties of 0.09 mag. For those objects observed on both
1998 Dec 3 and on 1998 Dec 8, the two measurements agree within the 
uncertainties. In what follows the mean of each pair of observations is reported.

Table~\ref{table2} summarizes the new $JHKL'MN$ measurements. As can
be seen from this Table, we obtained $N$-band detections among only 
six of the 17 targets we observed. 
For the remainder of the targets observed at $N$, we were able to measure
sensitive flux upper limits. As we discuss
in \S 3, these data allow us to place meaningful limits on
the presence of disks surrounding these objects. 

\begin{deluxetable}{lrrrrrrrrrrrrrrrrrl}
\tabletypesize{\scriptsize}
\tablecolumns{19}
\tablecaption{Broadband Photometry and Derived Stellar Parameters for Study 
Sample\label{table2}}
\tablehead{
\colhead{Star} & \colhead{$P_{\rm rot}$} & \colhead{$U$} & \colhead{$B$} & \colhead{$V$} 
& \colhead{$R$}\tablenotemark{1} &
\colhead{$I_C$} & \colhead{$J$} & \colhead{$H$} & \colhead{$K$} & \colhead{$L$,$L'$} &
\colhead{$M$} & \colhead{$N$} & \colhead{$T_{\rm eff}$} & \colhead{$L_\star$} &
\colhead{$M_\star$} & \colhead{$R_\star$} & \colhead{$A_V$} & \colhead{Refs.} \\
\colhead{} & \colhead{days} & \colhead{mag} & \colhead{mag} & \colhead{mag} & \colhead{mag} &
\colhead{mag} & \colhead{mag} & \colhead{mag} & \colhead{mag} & \colhead{mag} &
\colhead{mag} & \colhead{mJy} & \colhead{K} & \colhead{${\rm L}_\sun$} &
\colhead{${\rm M}_\sun$} & \colhead{${\rm R}_\sun$} & \colhead{mag} & \colhead{} }
\startdata
\cutinhead{Targets with new data}
\sidehead{ONC stars}
\objectname[]{JW 330} & 1.57 & \nodata & \nodata & 13.20 & \nodata & 11.78 & 10.77 &
   10.05 & 9.87 & \nodata & \nodata & {\it 17.4}\phs\phn\phn\phd\phn & 4775 & 2.82 & 1.07 & 2.46 & 0.85 & 1,3,15 \\
\objectname[]{JW 669} & 1.81 & \nodata & \nodata & 12.43 & \nodata & 10.76 & 9.43 & 8.73 & 8.52 &
   \nodata & \nodata & {\it 36.2}\phs\phn\phn\phd\phn & 5105 & 13.18 & 2.13 & 4.66 & 1.91 & 1,3,4,15 \\
\objectname[]{JW 794} & 2.62 & \nodata & \nodata & 11.67 & \nodata & 10.86 & 10.27 & 9.83 & 9.78 &
   \nodata & \nodata & {\it 13.8}\phs\phn\phn\phd\phn & 5272 & 4.17 & 1.93 & 2.46 & 0.00 & 1,3,15 \\
\objectname[]{JW 790} & 2.74 & 13.21 & 12.77 & 11.94 & {\it 11.12} & 10.88 & 10.06 & 9.55 & 9.44 &
   9.52 & \nodata & 20.9$\pm${\phn}7.7 & 5236 & 5.37 & 2.04 & 2.83 & 0.49 & 1,3,10,15 \\
\objectname[]{JW 641} & 3.17 & \nodata & \nodata & 13.25 & \nodata & 11.51 & 10.27 & 9.46 & 9.27 &
   \nodata & \nodata & {\it 16.9}\phs\phn\phn\phd\phn & 5236 & 8.13 & 2.24 & 3.46 & 2.24 & 1,3,15 \\
\objectname[]{JW 544} & 3.42 & \nodata & \nodata & 13.04 & \nodata & 10.82 & 9.35 & 8.45 & 8.23 &
   \nodata & \nodata & {\it 29.6}\phs\phn\phn\phd\phn & 4677 & 19.95 & 1.19 & 6.84 & 2.76 & 1,3,15 \\
\objectname[]{JW 926} & 5.59 & \nodata & \nodata & 14.68 & \nodata & 12.75 & \nodata & \nodata & 10.53 &
   \nodata & \nodata & 27.0$\pm${\phn}8.7 & 4000 & 1.29 & 0.40 & 2.37 & 0.88 & 2,3,4,15 \\
\objectname[]{JW 866} & 6.76 & \nodata & \nodata & 13.95 & \nodata & 11.82 & 10.10 & 9.25 & 8.95 &
   \nodata & \nodata & {\it 26.1}\phs\phn\phn\phd\phn & 5105 & 9.77 & 1.95 & 3.98 & 3.09 & 1,3,4,15 \\
\objectname[]{JW 352} & 8.00 & \nodata & \nodata & 12.46 & \nodata & 10.36 & 8.90 & 8.12 & 7.84 &
   7.58 & 7.98 & {\it 25.9}\phs\phn\phn\phd\phn & 3890 & 12.30 & 0.35 & 7.68 & 0.99 & 1,3,4,15 \\
\objectname[]{JW 567} & 8.53 & \nodata & \nodata & 11.39 & \nodata & 9.66 & 8.73 & 7.91 & 7.66 &
   7.20 & 7.58 & \nodata\phs\phn\phn\phd\phn & 4581 & 26.92 & 1.04 & 8.23 & 1.36 & 1,3,4,15 \\
\objectname[]{JW 589}\tablenotemark{2} & 14.3\phn & \nodata & \nodata & 13.42 & \nodata & 11.30 & 9.63 & 
   8.75 & 8.53 & \nodata & \nodata & {\it 19.0}\phs\phn\phn\phd\phn & 5236 & 16.98 & 2.79 & 5.01 & 3.21 & 1,3,15 \\
\objectname[]{JW 157} & 17.4\phn & 13.60 & 12.83 & 11.71 & {\it 10.58} & 10.15 & 8.94 & 8.24 & 8.06 &
   \nodata & \nodata & 33.8$\pm${\phn}6.8 & 4775 & 15.51 & 1.34 & 5.76 & 1.21 & 1,3,11,15 \\
\sidehead{Tau-Aur stars}
\objectname[]{TAP 41} & 2.43 & 14.30 & 13.30 & 12.06 & 11.31 & 10.61 & 9.59 & 8.92 & 8.77 & 
   8.73 & 8.74 & {\it 20.7}\phs\phn\phn\phd\phn & 4060 & 0.65 & 0.7\phn & 1.65 & 0.00 & 5,6,12,15 \\
\objectname[]{V830 Tau} & 2.75 & 14.71 & 13.61 & 12.23 & 11.37 & 10.51 & 9.45 & 8.72 & 8.55 & 
   8.28 & \nodata & 21.7$\pm${\phn}8.0 & 4060 & 0.89 & 0.7\phn & 1.93 & 0.28 & 5,6,7,12,15 \\
\objectname[]{V826 Tau}\tablenotemark{b} & 3.7\phn & 14.62 & 13.50 & 12.11 & 11.23 & 10.33 & 9.07 & 8.35 & 8.19 &
   8.07 & \nodata & 28.3$\pm$10.7 & 4060 & 0.89 & 0.5\phn & 1.93 & 0.28 & 5,6,7,12,15 \\
\objectname[]{IW Tau}\tablenotemark{b} & 5.6\phn & 15.23 & 14.04 & 12.52 & 11.50 & 10.60 & 9.16 & 8.40 & 8.20 &
   8.15 & \nodata & {\it 19.8}\phs\phn\phn\phd\phn & 4060 & 0.87 & 0.7\phn & 1.90 & 0.83 & 5,6,7,12,15 \\
\objectname[]{TAP 45} & 6.20 & 16.06 & 14.72 & 13.22 & 12.28 & 11.37 & 10.10 & 9.40 & 9.29 &
   9.18 & \nodata & 12.8$\pm${\phn}5.9 & 4060 & 0.35 & 0.7\phn & 1.21 & 0.69 & 5,6,12,13,15 \\
\cutinhead{Additional Tau-Aur targets from the literature}
\objectname[]{HP Tau/G2} & 1.20 & 13.38 & 12.46 & 11.07 & 10.21 & 9.36 & 8.10 & 7.38 & 7.19 &
   6.84 & 6.76 & 83.4\phs\phn\phn\phd\phn & 6030 & 10.40 & 1.9\phn & 2.98 & 2.08 & 5,6,7 \\
\objectname[]{HD 283572} & 1.55 & 10.19 & 9.87 & 9.04 & 8.56 & 8.10 & 7.46 & 7.04 & 6.93 &
   6.85 & \nodata & 83.6$\pm${\phn}6.6 & 5770 & 10.81 & 2.0\phn & 3.32 & 0.38 & 5,6,7,8 \\
\objectname[]{LkCa 19} & 2.24 & 12.48 & 11.87 & 10.85 & 10.25 & 9.68 & 8.85 & 8.27 & 8.15 &
   8.05 & \nodata & {\it 5.4}\phs\phn\phn\phd\phn & 5250 & 1.50 & 1.4\phn & 1.50 & 0.00 & 5,6,7,8 \\
\objectname[]{LkCa 1} & 2.50 & 16.26 & 15.22 & 13.73 & 12.52 & 11.07 & 9.65 & 8.91 & 8.66 & 
   8.53 & \nodata & 22.6$\pm${\phn}6.8 & 3370 & 0.77 & 0.25 & 2.60 & 0.00 & 6,7,8,14 \\
\objectname[]{TAP 35} & 2.74 & 11.44 & 11.13 & 10.34 & 9.88 & 9.45 & 8.90 & 8.49 & 8.40 &
   8.24 & \nodata & 19.5$\pm${\phn}3.9 & 5080 & 1.81 & 1.3\phn & 1.75 & 0.10 & 5,6,7,8,12 \\
\objectname[]{LkCa 4} & 3.37 & 15.16 & 13.96 & 12.49 & 11.54 & 10.56 & 9.28 & 8.57 & 8.35 &
   8.17 & \nodata & 20.0$\pm${\phn}5.0 & 4060 & 1.11 & 0.51 & 2.15 & 0.69 & 5,6,7,8 \\
\objectname[]{V773 Tau}\tablenotemark{b} & 3.43 & 13.12 & 12.02 & 10.65 & 9.80 & 8.94 & 7.63 & 6.83 & 6.48 &
   5.86 & \nodata & 2500\tablenotemark{a} & 4730 & 9.1\phn & 1.0\phn & 4.54 & 1.32 & 5,6,7,8 \\
\objectname[]{V827 Tau} & 3.75 & 14.70 & 13.58 & 12.18 & 11.29 & 10.34 & 9.05 & 8.33 & 8.13 &
   7.97 & \nodata & 28.9$\pm${\phn}4.3 & 4060 & 1.11 & 0.5\phn & 2.15 & 0.28 & 5,6,7,8,12 \\
\objectname[]{V819 Tau}\tablenotemark{b} & 5.6\phn & 16.12 & 14.81 & 13.24 & 12.24 & 11.16 & 9.59 & 8.77 & 8.50 &
   8.29 & \nodata & 28.5$\pm${\phn}4.2 & 4060 & 0.75 & 0.5\phn & 1.77 & 1.35 & 5,6,7,8,12 \\
\objectname[]{LkCa 7} & 5.64 & 15.11 & 13.94 & 12.55 & 11.63 & 10.58 & 9.26 & 8.56 & 8.33 &
   8.28 & \nodata & 24.8$\pm${\phn}5.2 & 4060 & 0.93 & 0.51 & 1.97 & 0.59 & 5,6,7,8 \\
\objectname[]{Anon 1} & 6.49 & \nodata & 15.37 & 13.52 & \nodata & \nodata & 9.01 & 7.88 & 7.50 &
   7.36 & \nodata & 39.2\phs\phn\phn\phd\phn & 3680\tablenotemark{3} & 1.30 & 0.38 & 2.59 & 1.32 & 6,7,14 \\
\objectname[]{LkCa 3}\tablenotemark{b} & 7.2\phn & 14.89 & 13.61 & 12.10 & 11.01 & 9.78 & 8.47 & 7.76 & 7.53 &
   7.32 & \nodata & 46.7$\pm$10.4 & 3720 & 1.66 & 0.29 & 3.13 & 0.42 & 5,6,7,8 \\
\objectname[]{DI Tau}\tablenotemark{b} & 7.5\phn & 15.96 & 14.46 & 12.86 & 11.81 & 10.74 & 9.41 & 8.63 & 8.40 &
   8.05 & \nodata & 100\tablenotemark{a} & 3850 & 0.62 & 0.4\phn & 1.79 & 0.76 & 5,6,7 \\
 &  &  &  &  &  &  &  &  &  &  &  & 22.8$\pm${\phn}6.8\tablenotemark{c} &  &  &  &  &  &  \\
\objectname[]{TAP 57NW} & 9.3\phn & 13.97 & 12.88 & 11.60 & 10.81 & 10.05 & 9.21 & 8.46 & 8.27 &
   8.07 & \nodata & 33.2$\pm${\phn}4.8 & 4060 & 1.09 & 0.8\phn & 2.13 & 0.00 & 5,6,7,8,12,13 \\
\objectname[]{GM Aur} & 12.0\phn & 13.55 & 13.22 & 12.03 & 11.22 & 10.50 & 9.37 & 8.73 & 8.48 &
   8.22 & \nodata & 190\tablenotemark{a} & 3950 & 0.83 & 1.0\phn & 1.37 & 0.14 & 5,6,7,8 \\
\enddata
\tablenotetext{1}{$R$ magnitudes are given in the Cousins system. Italicized values are
given in the Johnson system.}
\tablenotetext{2}{Very bright 10$\mu$m source in reference beam, probably \objectname[]{JW 548}.}
\tablenotetext{3}{Adopting this $T_{\rm eff}$ (corresponding to an M1 spectral type) instead
of the value published in \citet{kenyonhartmann} (corresponding to an M0 spectral type) provides
a better match to the observed optical SED.}
\tablenotetext{a}{Estimated from IRAS 12 $\mu$m flux \citep{skrutskie}.}
\tablenotetext{b}{Listed as a binary in \citet{kenyonhartmann}.}
\tablenotetext{c}{\objectname[]{DI Tau} was previously unresolved from the nearby 
\objectname[]{DH Tau} at these wavelengths. \citet{meyer97b} have
presented a revised, spatially resolved, $N$-band measurement for 
\objectname[]{DI Tau}. Also see text (\S 3.2) for
further discussion of this object.}
\tablerefs{
(1) \citet{herbst99},
(2) \citet{stass},
(3) \citet{hill97},
(4) \citet{hill98},
(5) \citet{bouvier95},
(6) \citet{kenyonhartmann},
(7) \citet{strom89},
(8) \citet{skrutskie},
(9) \citet{hartigan90},
(10) \citet{penston75},
(11) \citet{penston73},
(12) \citet{walter88},
(13) \citet{wolkwalter},
(14) \citet{grankin},
(15) this study }
\tablecomments{$N$-band measurements in italics are $3\sigma$ upper limits.}
\end{deluxetable}

\subsection{Data from the literature}
We supplement our new photometric measurements
with visible (i.e.\ $UBVRI$) to far-IR (i.e.\ {\it IRAS}) photometry from 
the literature to form SEDs that are as complete as possible in 
wavelength coverage. We use stellar parameters derived from optical
spectroscopy and photometry---visual extinction 
($A_V$), effective temperature ($T_{\rm eff}$),
luminosity ($L_\star$), mass ($M_\star$), radius ($R_\star$)---to
generate model SEDs for comparison to the observed SEDs.

For ONC targets listed in Table~\ref{table1}, we draw from the database
of visible photometry and spectroscopy compiled by \citet{hill97},
which gives apparent $V$ and $I$ magnitudes, $A_V$, $T_{\rm eff}$, $L_\star$,
$M_\star$, and $R_\star$. For those ONC targets which we did not observe
in $JHK$, we instead use the $JHK$ photometry of \citet{hill98}. 
For Tau-Aur targets listed in Table~\ref{table1}, we take broadband
$UBVRIJHK$ (and in some cases $LM$) photometry, {\it IRAS} fluxes, and
derived stellar parameters, primarily from the extensive databases
compiled by \citet{kenyonhartmann} and \citet{strom89}. The
compiled $UBVRI$ measurements are on both the Johnson and Cousins
photometric systems. While the compiled ONC $JHK$ measurements are
all on the CIT photometric system, some of the Tau-Aur $JHK$
measurements are on different systems for which the effective
wavelengths differ from the CIT effective wavelengths, but only slightly
\citep{bessell}. Throughout this paper,
we reference fluxes to the absolute system adopted by \citet{rydgren84},
with the exception that for $N$-band measurements obtained with the
OSCIR instrument, we use the zero-point flux of 
37.77 Jy adopted by the OSCIR team (S. Fisher, priv. comm.), which is
approximately 5\% larger than that used by \citet{rydgren84}.

In Table~\ref{table2} we reproduce the broadband magnitudes and derived
stellar parameters from the literature for targets listed in Table~\ref{table1}.
In addition, Table~\ref{table2} lists broadband magnitudes and 
stellar parameters for an additional 15 targets in Tau-Aur for which rotation
periods have been determined, which show near-IR colors consistent with
photospheric colors, and for which $N$-band fluxes exist in the 
literature. The visible and near-IR measurements discussed in 
\citet{kenyonhartmann} and in \citet{strom89} are characterized by
small photometric errors, typically $\sim 0.05$ mag. We do not, therefore,
show the uncertainties at these wavelengths in our compiled SEDs. 
However, we do show the $N$-band uncertainties reported in the 
literature, when provided.

Together, the 32 targets listed in Table~\ref{table2}
form the sample for this study. These stars are roughly equally divided
between ``rapid" ($P_{\rm rot} < 4$ days) and ``slow" ($P_{\rm rot} > 4$
days) rotators, as defined by previous authors \citep{edwards93,ch,herbst99},
and possess a range of rotation periods representative of TTS generally, with
$1.5 < P_{\rm rot} < 17$ days. 

\subsection{SED modeling procedure}
To compare the SEDs observed among stars in our sample to those expected 
from stars with truncated
circumstellar disks, we generate model SEDs following the procedure 
used by \citet{jensen} in their investigation of 
cavities in disks around PMS binaries. 
We fix various model parameters for each star using values 
for $A_V$, $T_{\rm eff}$, and $L_\star$ from the literature (see Table 2). 
The inner truncation radius of the disk, $R_{\rm trunc}$, is 
set equal to the co-rotation radius, $R_{\rm c}$, computed from
the observed stellar rotation period, $P_{\rm rot}$, and stellar
mass, $M_\star$ (see Table 2). For each star, we generate models with 
disk inclinations, $i$, of $0^\circ$ (i.e. face-on) and $60^\circ$.
We normalize the model photospheric flux to the observed $I$-band flux.

For each star in our sample, we calculate the emission from a disk
that is optically thick at visible and infrared wavelengths and is
continuous from $R_{\rm trunc}$ to an outer radius of 100 AU. 
We assume that the disk has no source of energy other than the
reprocessing of stellar photons. 
Such a disk has the {\it minimum} luminosity possible for an
optically thick disk; the disk cannot be 
less luminous unless it becomes optically thin.
Thus, if the observed emission from a star at a given wavelength
falls below this model, the implication is that the disk is absent
(or at least optically thin) in the
regions that dominate the disk emission at that wavelength. 
Since our models have $R_{\rm trunc} = R_{\rm c}$,
if the observed emission falls below the model SED
we infer that disk material is absent out to {\it at least}
the co-rotation distance from the star.

\section{Results}
In Figures \ref{oncsedfig} and \ref{tausedfig} we compare the observed
SEDs of stars in our sample (from Table 2) to model star$+$disk SEDs. 
For each star, the model SED consists of a single star with the same
$T_{\rm eff}$ as the target surrounded by a flat, reprocessing disk 
with $R_{\rm trunc} = R_{\rm c}$. 
We consider the evidence for truncated disks in turn among ONC stars
and Tau-Aur stars in our sample.

\subsection{ONC stars}
Among the 12 ONC stars in our study sample, only one star 
(\objectname[]{JW 926}) shows evidence for excess mid-IR 
emission (Fig. \ref{oncsedfig}), albeit marginal (3$\sigma$).
The observed $N$-band flux for this star, in light of the observed 
$K$-band flux being photospheric, is
consistent with the model SED prediction for a truncated disk 
with $R_{\rm trunc} \approx R_{\rm c}$. This star is a moderately
slow rotator with $P_{\rm rot} = 5.6$ days.

None of the remaining 11 ONC stars in our study sample 
show evidence for significant excess mid-IR emission. None of the 
stars with $N$-band detections show fluxes significantly ($>2\sigma$)
in excess of photospheric,
and all stars, even those for which only upper limits are available, 
indicate that no disk material is present at or near $R_{\rm c}$.
Indeed, the observed $N$-band detections and upper limits are so
far below the predicted model SED fluxes that, if optically thick disks
are present around these stars, they must have $R_{\rm trunc} \gg
R_{\rm c}$; for most of the stars in our sample,
$R_{\rm trunc} \gtrsim 1$ AU would be required to be consistent with
photospheric $N$-band fluxes. 
We note that two stars in our sample 
(\objectname[]{JW 567} and \objectname[]{JW 589})
show weak evidence for active accretion in the form of marginally
detected \ion{Ca}{2} triplet emission \citep{hill98}, but the
newly determined SEDs for these stars (Fig.\ \ref{oncsedfig}) fairly 
conclusively rule out the existence of (optically thick) 
inner disk material.

To emphasize the extent to which the ONC stars in our sample are 
lacking in inner
disk material, we show in Figure \ref{bigholefig} the effect of
increasing $R_{\rm trunc}$ beyond $R_{\rm c}$ for the star 
\objectname[]{JW 352}, which is a ``typical" ONC slow rotator with
$P_{\rm rot} = 8.0$ days and $M_\star = 0.35 {\rm M}_\odot$ 
($R_{\rm c} = 0.07$ AU). The
figure clearly illustrates that, even though the model SED assumes
the minimum disk luminosity possible, an extremely large inner disk
hole ($R_{\rm trunc} \approx 1.5$ AU) is required to eliminate the
disk flux at $N$. Thus, insofar as we are concerned with disks that
may be acting to regulate stellar rotation, the ONC stars in our
sample are effectively diskless.

\subsection{Tau-Aur stars}
We derive similar results from the Tau-Aur stars in our study sample
(Fig. \ref{tausedfig}). Three stars in our sample (\objectname[]{V773 Tau},
\objectname[]{DI Tau}, \objectname[]{GM Aur}) show evidence for 
significant excess mid-IR emission. Considering these stars' photospheric
near-IR fluxes, this mid-IR emission is roughly consistent
with that expected from disks with $R_{\rm trunc} \approx R_{\rm c}$. 
Interestingly, these objects have associated peculiarities that
could modify the simple interpretation of their SEDs as arising from
truncated disks. \objectname[]{V773 Tau} is a triple system including
a close binary \citep{ghez}, making a clear partitioning of the system's mid-IR 
flux difficult. \objectname[]{GM Aur} has been considered in detail by 
\citet{koerner93} as a candidate for inner-disk ``clearing" by a
sub-stellar, perhaps planetary, companion. \objectname[]{DI Tau} 
has been closely investigated by \citet{meyer97b}, who found this star's
10 $\mu$m flux to be consistent with being purely photospheric, its 
previously measured 10 $\mu$m excess in fact being due to the 
nearby object \objectname[]{DH Tau}.
In what follows, we thus take \objectname[]{DI Tau} to be diskless,
but continue to regard \objectname[]{V773 Tau} and
\objectname[]{GM Aur} as possible systems with truncated disks, keeping
the above caveats in mind.
\objectname[]{V773 Tau} is a rapid rotator ($P_{\rm rot} = 3.4$ days), 
while \objectname[]{GM Aur} is a slow rotator (12.0 days).

As with the ONC stars in our sample, the remaining 17 
Tau-Aur stars in our sample lack significant excess mid-IR emission.
Combining these results with those of our ONC sample, we find that 
29/32 TTS lacking
near-IR excess-emission signatures of disks show SEDs consistent with
being purely photospheric for $\lambda \le 10$ $\mu$m (and in some cases 
for $\lambda > 10$ $\mu$m). 
The general tendency for these stars to exhibit only photospheric-level
$N$-band fluxes indicates the absence of disk material at (and, indeed,
well beyond) $R_{\rm c}$; disks are not presently in a position to 
regulate the rotation of these stars.

\section{Discussion}
In the previous section we used new and existing mid-IR photometry to 
search for truncated disks among 32 TTS known to lack excess emission 
at near-IR wavelengths. The basic result of this study
is that, with a few possible exceptions, the young PMS stars that 
comprise our sample---both rapid and slow rotators---lack circumstellar 
disks that might act to regulate their rotation.
In this section, we consider the implications of this result
for the general question
of the role that disks play in the regulation of 
stellar rotation among low-mass PMS stars. We begin by comparing
our findings with those of previous studies that have
linked the presence of disks to slow rotation in TTS. 
Next, we discuss the implications of our
results for rotational evolution models of low-mass PMS stars.
Finally, we consider the extent to which our observations
and those of other authors agree with the predictions of magnetic
disk-locking models. 

\subsection{Disked slow rotators and diskless rapid rotators: A false 
dichotomy?}
As one of the pillars supporting the idea of disk-regulated stellar
angular momentum, several authors have linked slow stellar rotation 
to IR signatures of disks \citep{edwards93,bouvier93,ch,herbst99}.
Within this context, slowly rotating TTS ($P_{\rm rot} > 4$ days)
are understood in terms of magnetic locking to disks, while rapidly 
rotating TTS ($P_{\rm rot} < 4$ days) are interpreted as
stars that were disk-locked in an earlier epoch, but that have spun 
up after having lost their disks. Though
slow and rapid rotators in the ONC and in Taurus-Auriga 
in fact appear roughly coeval in terms of their
placement in the H-R diagram \citep{stass}, 
this separation of TTS into slow rotators with disks and rapid
rotators without disks has resulted in slowly and rapidly rotating TTS 
being seen in terms of an {\it evolutionary sequence}, one
in which rapid rotators {\it derive from} 
slow rotators. Consequently, slow rotators have
come to define the ``initial conditions" assumed by most PMS rotational 
evolution models \citep{keppens95,bouvier97,krishnamurthi};
rapidly rotating TTS have in general not been considered as part of the 
initial conditions for the models.

However, while the \citet{edwards93} study found no instances of 
rapidly rotating TTS with significant near-IR signatures of disks,
more recent studies have in fact found disked rapid rotators,
making the connection between disks and slow rotators
less distinct. For example, \citet{eaton} found significant
excess near-IR emission among two of the five rapid rotators 
in their study of the Trapezium. In our recent study of the ONC 
(Paper~I), involving a considerably larger sample,
we found roughly equal numbers of rapid rotators with and without strong 
near-IR excesses; and the same is true of the similarly sized
ONC sample considered by \citet{herbst99}.
We also found in Paper~I that 
strong H$\alpha$ emission, indicative of active disk accretion, is present 
in rapid rotators with roughly equal frequency to the slow rotators in 
our sample. Thus, it now seems that there exist large numbers of
rapidly rotating TTS with circumstellar disks. Indeed, the fraction
of rapid rotators with near-IR signatures of disks observed by us in 
Paper~I is roughly consistent with the frequency of disks generally among 
low-mass stars in the ONC \citep{hill98}.

In a similar vein, a substantial fraction of slow rotators
evidently lack disks altogether.  Roughly one-third of the
slowly rotating TTS in the sample studied by \citet{edwards93} 
were observed to lack excess near-IR emission.
More recent studies \citep{stass,herbst99} have found a similar 
fraction of slow rotators without strong near-IR excesses. 
Guided by the predictions of magnetic disk-locking models, some authors 
have therefore hypothesized that slow rotators lacking near-IR excesses may in 
fact possess disks, but with large truncation radii.
\citet{edwards93}, for example,
cited two slow rotators in that study possessing mid-IR excesses
(\objectname[]{DI Tau}\footnote{New spatially resolved 
mid-IR photometry presented by \citet{meyer97b} reassigns the excess
mid-IR emission previously attributed to \objectname[]{DI Tau} to
\objectname[]{DH Tau}. See \S 3.2.} and \objectname[]{GM Aur}) 
as possible examples of this phenomenon. 
However, the results of this study indicate that, generally speaking,
{\it slowly rotating TTS lacking near-IR excesses simply do not possess disks
truncated at} $\approx R_c$.

Thus, a division of
TTS into disked slow rotators and diskless rapid rotators
is not an accurate
representation of the observed relationship between stellar rotation
and the presence of IR excesses among low-mass PMS stars at 1 Myr.
While there is evidence that slow rotators tend to exhibit larger near-IR 
excesses than do rapid rotators \citep{herbst99},
our observations do not support the 
interpretation that rapid rotators are, as a group, significantly different than slow 
rotators with respect to disk frequency. 
This result is important not only because it questions the idea
of a linkage between disks and slow stellar rotation, but because it 
challenges the observational basis for the idea that the rapid rotators 
present at 1 Myr are the evolutionary descendants of once-slow rotators.
This has important implications for models of PMS rotational evolution,
as we now discuss.

\subsection{Implications for models of PMS rotational evolution}
Current rotational evolution models are able to
reproduce in some detail the observed evolution of rotation rates 
of low-mass stars
over a range of ages, from about 1 Myr to the present-day Sun. 
Such a statement represents a triumph at both extremes of stellar 
rotation; a long-standing problem in modeling the rotational 
evolution of low-mass stars has been to simultaneously explain 
the origin of both slow rotators 
($v\sin i \lesssim 10$ km/s) and ultra-fast rotators 
($v\sin i \sim 150$ km/s) on the ZAMS. 
The nature of this ``slow and
rapid rotator problem" has been discussed in detail in several
recent studies \citep{barnes96,bouvier97,krishnamurthi,barnes99}.
In essence, this problem has been one of dispersion. 
The question faced by rotational evolution models has been:
By what means is the large dispersion of rotation rates 
observed on the ZAMS created? 

Implicit in this question is the assumed {\it model initial conditions},
which are a fundamental ingredient in constructing 
the rotational evolution models. The initial conditions adopted
by most current rotational evolution models are predicated
upon previous observations of TTS in Taurus and Orion which
have linked the presence of disks to slow stellar rotation, and which
have suggested that slow rotators precede rapid rotators in an
evolutionary sequence. Thus, the models typically begin their 
calculations with a population of
slow rotators (i.e. $P_{\rm rot} \sim 8$ days) possessing a relatively 
small dispersion of rotation rates \citep{bouvier97,krishnamurthi}. 

However, the evidence presented in this paper and in Paper~I 
questions some of the assumptions that have guided the choice
of model initial conditions. 
As we discussed in the previous section, we have not found compelling
evidence on the basis of disk frequency that the rapid rotators present
at 1 Myr collectively represent an evolved state of the slow rotators.
Rather, our observations 
suggest that {\it the rapid rotators should be considered just as important
as the slow rotators for defining the model initial conditions}. 

Considering the most rapidly rotating TTS ($P_{\rm rot} \sim 0.5$ days; Paper~I)
together with the slow
rotators gives the low-mass PMS population at 1 Myr a very large dispersion 
of rotation rates, one that matches---or even exceeds---that 
observed among low-mass Pleiads ($\sim 100$ Myr). 
While the differences observed among Pleiads in 
different mass ranges may suggest somewhat different rotational 
histories for stars of different masses \citep{terndrup00},
one thing is clear: {\it The dispersion of stellar rotation rates observed
among low-mass ZAMS stars is already present at 1 Myr} \citep{stass,ch}. 
The solution to the ``dispersion problem" of at once accounting for the slow rotators
and the UFRs on the ZAMS may largely depend upon taking proper account of the large 
dispersion that is already present at 1 Myr.

Whether the large dispersion of rotation rates present at 1 Myr is itself
a product of early angular momentum evolution (perhaps disk-regulated), 
or simply a reflection of the initial conditions of star formation, 
or some combination of the two, is a question that observations and rotational 
evolution models must now answer.
While observations of TTS are presently our primary informants of the initial
conditions of rotational evolution, TTS at 1 Myr do not themselves represent 
the ``beginning" of the 
angular momentum evolution problem. As \citet{herbst99} have emphasized, 
disk-locking may act to significantly impact the angular momentum of 
stars prior to even the very young age of the ONC. 
Eventually, rotational studies of still-embedded objects (via, e.g.,
photopolarimetric \citep{stasswood} and scattered-light modeling
\citep{woodwhitney} techniques, or high-resolution IR spectroscopy
\citep{greene}), and of still-collapsing
cloud cores, will be necessary to truly understand the 
initial conditions of angular momentum in young stars. Nonetheless, the
large dispersion of rotation rates present at 1 Myr represents a critical
early touchstone for rotational evolution models, and may prove to be a
significant challenge to models that do not include rapid rotators
at the beginning of their calculations \citep{sills}.

\subsection{The interpretation of near-IR excesses:
Evidence for disk-locking?}
As we have seen, young, low-mass PMS stars that do not 
show excess near-IR emission do not, in general, harbor truncated disks. 
Put another way, disks, when present, nearly always reveal 
themselves at near-IR wavelengths. What does this tell us about
the typical structure of TTS disks, and how does this
compare with the requirements of magnetic disk-locking models?

Numerous researchers have investigated the details of magnetic
coupling between an accreting TTS and its circumstellar disk
\citep[and others]{konigl91,yi94,ostrikershu,armitage96}. 
An important prediction of magnetic disk-locking models is that
the disk should truncate at a radius, $R_{\rm trunc}$, that is close 
to the co-rotation radius, $R_{\rm c}$, if the disk is to 
effectively regulate the star's rotation while at the same
time allowing for accretion onto the stellar surface. \citet{wang95}, 
for example, employs several lines of argument to demonstrate that
$R_{\rm trunc} / R_{\rm c} \approx 0.9$ is required in order for 
the magnetic star-disk interaction to prevent an accreting TTS
from experiencing significant spin-up torques. 

Is this basic prediction of the disk-locking hypothesis---that 
$R_{\rm trunc} \approx R_{\rm c}$---in fact
consistent with the near-IR observations of TTS?
Several authors have modeled the near-IR
emission from TTS disks, taking into account such factors
as inner disk truncation radius, disk accretion
rate, stellar rotation period, and stellar magnetic field 
strength \citep{kyh96,meyer97,armitage99}. \citet{kyh96} have also included the
effects of magnetic heating in their calculations.
A general result of these modeling efforts has been that 
disks with relatively small inner cavities---i.e.\ 
$R_{\rm trunc} \ll R_{\rm c}$---are required
to explain the observed near-IR emission of CTTS. 

As a simple demonstration of 
this point, we reproduce in Fig.\ \ref{montefig} 
a grid of SED models from \citet{thesis}, which are based on the SED 
modeling procedure of \citet{jensen}.
For these model calculations, 
$R_{\rm trunc}$ was varied from $R_\star$ (i.e. no inner disk hole) to 
10 $R_\star$, and the mass accretion rate through the disk, 
$\dot{M}$, was varied from 0 (i.e. a purely reprocessing
disk) to $10^{-6} {\rm M}_\odot {\rm yr}^{-1}$. 
In all models, the central star is characterized
by $T_{\rm eff} = 3900$K (appropriate for a spectral type
of $\sim$ M0) and $R_\star = 2 {\rm R}_\odot$, typical of
low-mass stars in the ONC \citep{hill97}. The models assume zero extinction,
and the disk inclination angle is fixed at $0^\circ$ (i.e. face on). 
For each model,
the near-IR ``excess-color" diagnostic used by us in
Paper~I and by \citet{herbst99}, $\Delta (I-K)$, is computed as the difference 
between the total star$+$disk $(I-K)$ color and the photospheric $(I-K)$ color.
The range of $R_{\rm trunc} / R_\star$ considered here corresponds to 
approximately $0.5 < P_{\rm rot} < 13$ days if $R_{\rm trunc} = R_{\rm c}$,
representative of the range of stellar rotation periods observed among TTS
\citep{bouvier95,ch,stass,herbst99}. 

Fig.\ \ref{montefig} illustrates that
disks with small $R_{\rm trunc}$ are needed to reproduce the 
very strong excesses [e.g.\ $\Delta (I-K) \gtrsim 1$ mag] that many
TTS exhibit \citep{herbst99}. Disk
models with large $R_{\rm trunc}$ produce only very modest excesses
[$\Delta (I-K) \lesssim 0.3$ mag]
for typical $\dot{M}$.
Thus disk-locked slow rotators with $R_{\rm trunc} \approx R_{\rm c}$
are {\it not} in general expected to show their disks in the near-IR. 
On the other hand, excess
emission at near-IR wavelengths {\it is} expected from stars that are 
disk-locked at relatively short rotation periods, for which
$R_{\rm trunc} \approx R_{\rm c}$ implies $R_{\rm trunc} \approx R_\star$. 
In other words, the 
correlation between near-IR excess and stellar rotation period
that has been reported in the literature \citep{edwards93,herbst99}---in
which the slowest rotators tend to exhibit the largest near-IR 
excesses---is in the {\it opposite sense} of that predicted from disks 
with $R_{\rm trunc} \approx R_{\rm c}$. 
This suggests that the condition required for the regulation of
stellar angular momentum---that the disk truncate near the co-rotation
radius---is {\it not} satisfied by most TTS disks (see also 
\citet{safier} who argues this point on theoretical grounds).

Interestingly, models in which the disk truncates well inside of 
co-rotation may be able to explain both the magnitude and the {\it sense} 
of the correlation between strength of near-IR excess and stellar 
rotation period observed by \citet{herbst99}. 
The basic idea here, as suggested by \citet{thesis}, rests
upon the fact that the most commonly used near-IR disk diagnostics---e.g.\ 
$\Delta (I-K)$---are really measures of 
``excess color", not excess flux per se. Consequently, 
disks with $R_{\rm trunc} \approx R_\star$
can show relatively small color-excesses that belie the presence of
substantial near-IR excess flux because a non-negligible disk contribution
at $I$ flattens the $(I-K)$ color. This 
effect becomes most pronounced for high $\dot{M}$ because as the inner
disk edge becomes hotter with increasing $\dot{M}$, the disk's
$(I-K)$ color becomes correspondingly bluer (e.g.\ Fig.\ \ref{montefig}). 
Over a limited range of $R_{\rm trunc}$, $\Delta (I-K)$ becomes larger
in magnitude as the inner edge of the disk becomes cooler, resulting
in a more rapid decline of the disk contribution at $I$ than at $K$. That is,
the disk's color becomes redder as the size of the central cavity is
increased \citep{meyer97}. But this effect does not hold for arbitrarily
large $R_{\rm trunc}$: If the
cavity is made too large, the disk contribution at $K$ is lost as well,
resulting in a loss of $(I-K)$ excess entirely. In this way, a correlation 
between near-IR ``excess color" and slow rotation might be produced 
even if $R_{\rm trunc}$ is typically some small fraction of $R_{\rm c}$. 

Magnetospheric accretion models in which $R_{\rm trunc} \ll R_{\rm c}$ 
have been considered previously
\citep{yi95,kyh96}.  In these models, the disk truncates at a point
that is determined by the balance between magnetospheric and accretion
pressures, relative to the co-rotation radius. In this way, $R_{\rm trunc}$
(and hence near-IR ``excess color") can correlate with $P_{\rm rot}$
(through the dependence of $R_{\rm c}$ on $P_{\rm rot}$), but
$R_{\rm trunc} / R_{\rm c}$ may deviate significantly from
unity depending upon $\dot{M}$ and $B_\star$. 
In fact, \citet{kyh96} calculate that $\dot{M} \sim 10^{-7}
{\rm M}_\odot {\rm yr}^{-1}$, $B_\star \sim {\rm few}\times 10^2$ G, 
and $P_{\rm rot} = 7$ days, results in 
$R_{\rm trunc} / R_{\rm c} \sim 0.4$--0.75, and less for
longer rotation periods. 
In this picture, 
the star-disk interaction may not in general be suitable for 
effecting disk-regulated stellar rotation in the manner usually 
envisioned in ``disk-locking" scenarios. 

These ideas are intriguing, but preliminary.
More fundamentally, we consider the question of disk truncation
as critical for any examination of the observational
evidence for disk-regulated stellar angular momentum, and
we encourage continued observational and modeling effort aimed
at determining whether TTS disks in fact truncate in the manner
required by disk-locking models.

\section{Summary and Conclusions}
We have used new and existing mid-IR photometry to search for circumstellar 
disks among 32 low-mass PMS stars in the ONC and in Taurus-Auriga
for which rotation periods have been measured
but which do not show evidence for disks at near-IR wavelengths. 
Our principal aim in this study has been to determine whether these
stars harbor disks with large inner truncation radii, or if they are simply 
diskless. The stars in our sample are roughly evenly divided between slow 
($P_{\rm rot} > 4$ days) and rapid ($P_{\rm rot} < 4$ days) rotators. 
The slow rotators, if shown to possess disks, would be consistent with
large truncation radii predicted by magnetic disk-locking models.
We similarly wish to determine if the rapid
rotators in our sample are coupled to truncated disks, perhaps undergoing
rapid spin-down at present, which 
could help explain the rotational fate of stars already rotating
near breakup velocity at 1 Myr \citep{stass}.
The most rapid rotators require a braking mechanism before reaching the
ZAMS; evidence for disk material might suggest non-steady-state
disk coupling.

The primary observational result of this study is that,
with a few possible exceptions, 
the stars in our sample---both the rapid and slow rotators---are not 
presently disk-locked, because they do not have disks to which the
stars may magnetically couple. Any disk material around these stars must
be situated at $\gtrsim 1$ AU, well beyond the typical co-rotation radius
of $\sim 0.05$ AU for the stars in our sample. 
By extension,
this implies that roughly one-third of slowly rotating 
PMS stars exist at 1 Myr without the aid of present disk-locking, and that 
a similar fraction of very rapidly rotating TTS (including 
some rotating near breakup velocity) will not have
their subsequent angular momentum evolution to the main sequence
regulated by circumstellar disks. 

Combining this new result with those of \citet{stass}, 
we draw two key interpretations with respect to magnetic disk-locking 
and rotational evolution models of low-mass PMS stars:
\begin{enumerate}
\item 
The dichotomy of disked slow rotators and diskless rapid rotators is a 
false one. Thus, we do not find compelling evidence on the basis of disk
frequency to suggest that slow rotators are the evolutionary 
precedents of rapid rotators at 1 Myr. 
Low-mass PMS stars at 1 Myr possess a large dispersion of rotation rates 
that is similar to---and that indeed may exceed---the 
dispersion observed among low-mass ZAMS stars. This includes a PMS
population of ``ultra-fast rotators" rotating near breakup velocity. 
The inclusion of rapid rotators at the beginning of rotational evolution
calculations may lessen the degree to which the models require the braking
action of disks \citep{sills}.

\item Disks, when present, do not appear to truncate at co-rotation as required 
by disk-locking models for the regulation of stellar angular momentum. 
Models of TTS disks suggest that disks must extend well 
inside of co-rotation, i.e. $R_{\rm trunc} \ll R_{\rm c}$, in order
to explain the near-IR excesses observed among disked TTS.
Furthermore, the previously
reported positive correlations between slow stellar rotation 
period and excess near-IR emission are in the
opposite sense of that predicted from disk-locking models. 
\end{enumerate}

These results, in concert with those previously discussed by us 
\citep{stass} and by others \citep{armitage99,safier,chakrabarty}, 
present some fundamental challenges to current models
of angular momentum evolution among young, low-mass PMS stars, which
have come to rely upon disk-locking.
The full impact of these conclusions awaits rotational evolution 
modeling that incorporates the rotational initial conditions advocated 
here, including rotational differences among PMS stars of different masses
\citep{herbst99}, as well as a more thorough investigation of 
angular momentum evolution prior to the stellar birthline.

\acknowledgments
It is a pleasure to thank H. Guetter for 
reducing the new USNO $JHK$ photometry. 
This research was supported by a Minority Dissertation Fellowship 
from the Ford Foundation, through the National Academy of Sciences 
(KGS), a grant from the Wisconsin Space Grant Consortium (KGS), and
grant AST-9417195 from the National Science Foundation (RDM).
Thesis observing support to Chile from NOAO is also gratefully acknowledged.
This work was improved by fruitful discussions with C. Dolan, S. Barnes,
and F. Shu.
Special thanks are due E. Jensen who generously provided us his SED modeling
code, and to the OSCIR team for their assistance in planning, obtaining,
and reducing the new 10 $\mu$m data presented here. This paper was improved by 
recommendations from an anonymous referee.

\section*{Figure captions}

\figcaption[]{SEDs for ONC stars in our sample. In
each panel, broadband photometric measurements (from Table 2) are plotted as 
filled squares, upper limits are plotted as arrows. The dashed line shows the 
model SED for the underlying photosphere, while the solid line shows the 
SED for the star$+$disk model with $R_{\rm trunc} = R_{\rm c}$. The dotted
line shows the same star$+$disk model with the disk inclined $60^\circ$.
The model photosphere has been normalized to the observed $I$-band flux.
For each star, the stellar rotation period (in days) is also indicated.
\label{oncsedfig}}

\figcaption[]{SED for JW 352 from Fig.\ \ref{oncsedfig} is
shown with three different star$+$disk SED models (solid lines), each
with a different value of $R_{\rm trunc}$: 0.1 AU, 0.5 AU, and 1.5 AU.
Each disk is inclined $60^\circ$.
Stars with photospheric $N$-band fluxes, especially slow rotators, must
have disks with $R_{\rm trunc} \gg R_{\rm c}$, if they possess disks at all.
\label{bigholefig}}

\figcaption[]{Same as Fig.\ \ref{oncsedfig}, except for Taurus-Auriga stars
in our sample.
\label{tausedfig}}

\figcaption[]{SED models from \citet{thesis} showing
the dependence of the near-IR ``excess-color" disk diagnostic,
$\Delta (I-K)$, as a function of $R_{\rm trunc} / R_\star$
for various values of the accretion rate, $\dot{M}$: $0$, $10^{-8}$,
$3\times 10^{-8}$, $10^{-7}$, $3\times 10^{-7}$, and $10^{-6}$,
in units of ${\rm M}_\odot {\rm yr}^{-1}$. The
highest curve in the plot corresponds to 
$\dot{M}=10^{-6} {\rm M}_\odot {\rm yr}^{-1}$.
The range of $R_{\rm trunc} / R_\star$ shown corresponds to about 
$0.5 < P_{\rm rot} < 13$ days when $R_{\rm trunc} = R_{\rm c}$.
The central star in the models has $T_{\rm eff} = 3900$K ($\sim$ M0) 
and $R_\star = 2 {\rm R}_\odot$. 
The horizontal line indicates the value of $\Delta(I-K)$
above which an observed excess would be considered significant.
\label{montefig}}

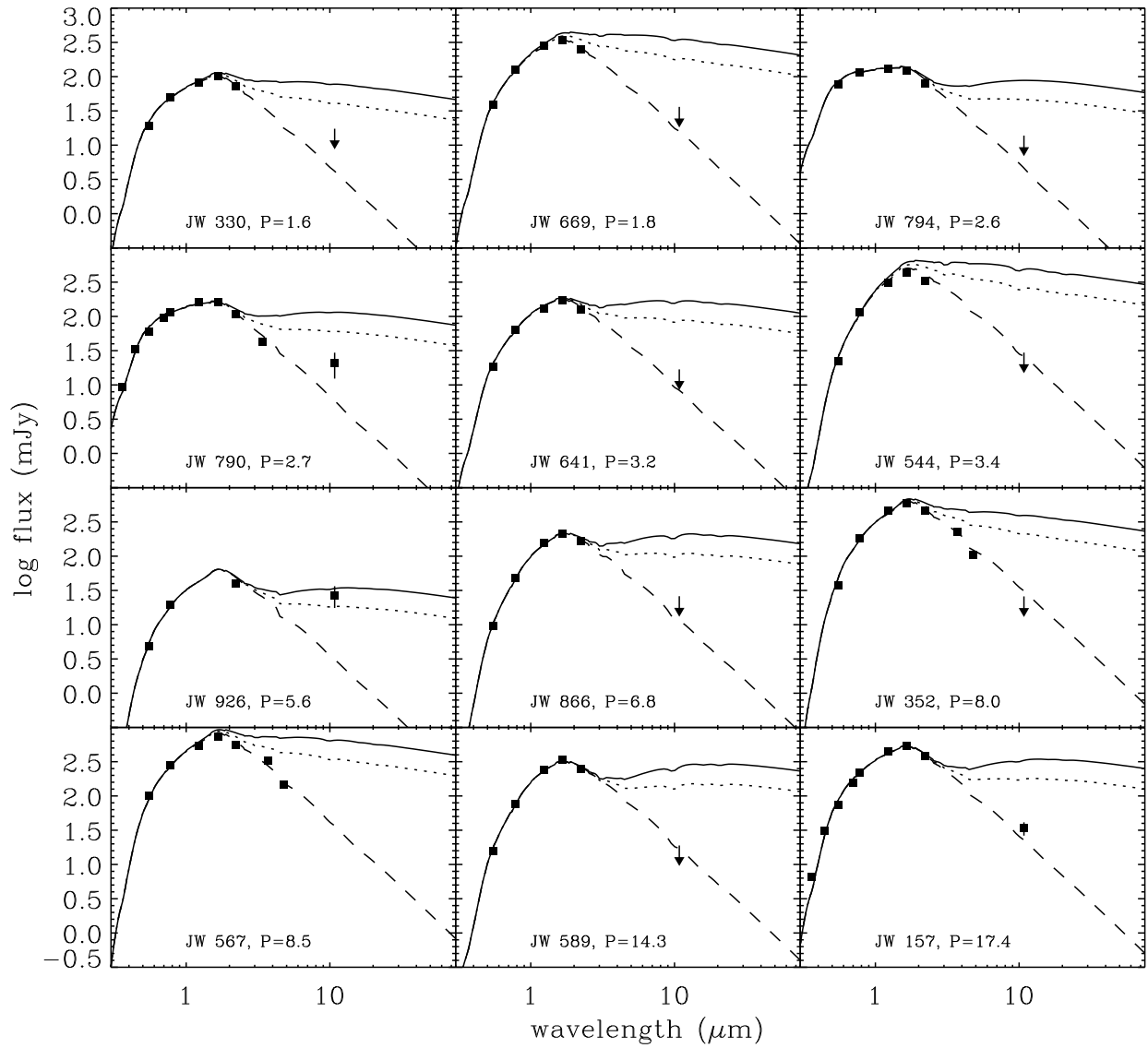
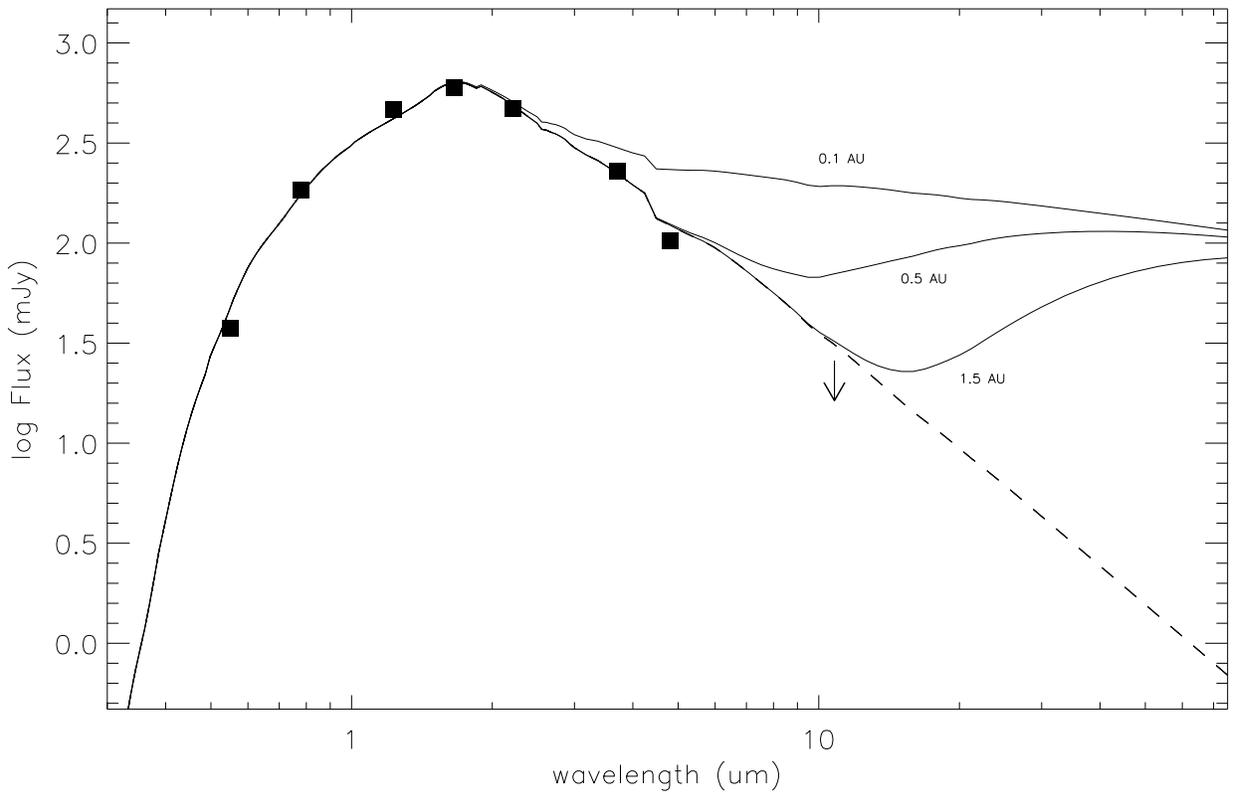
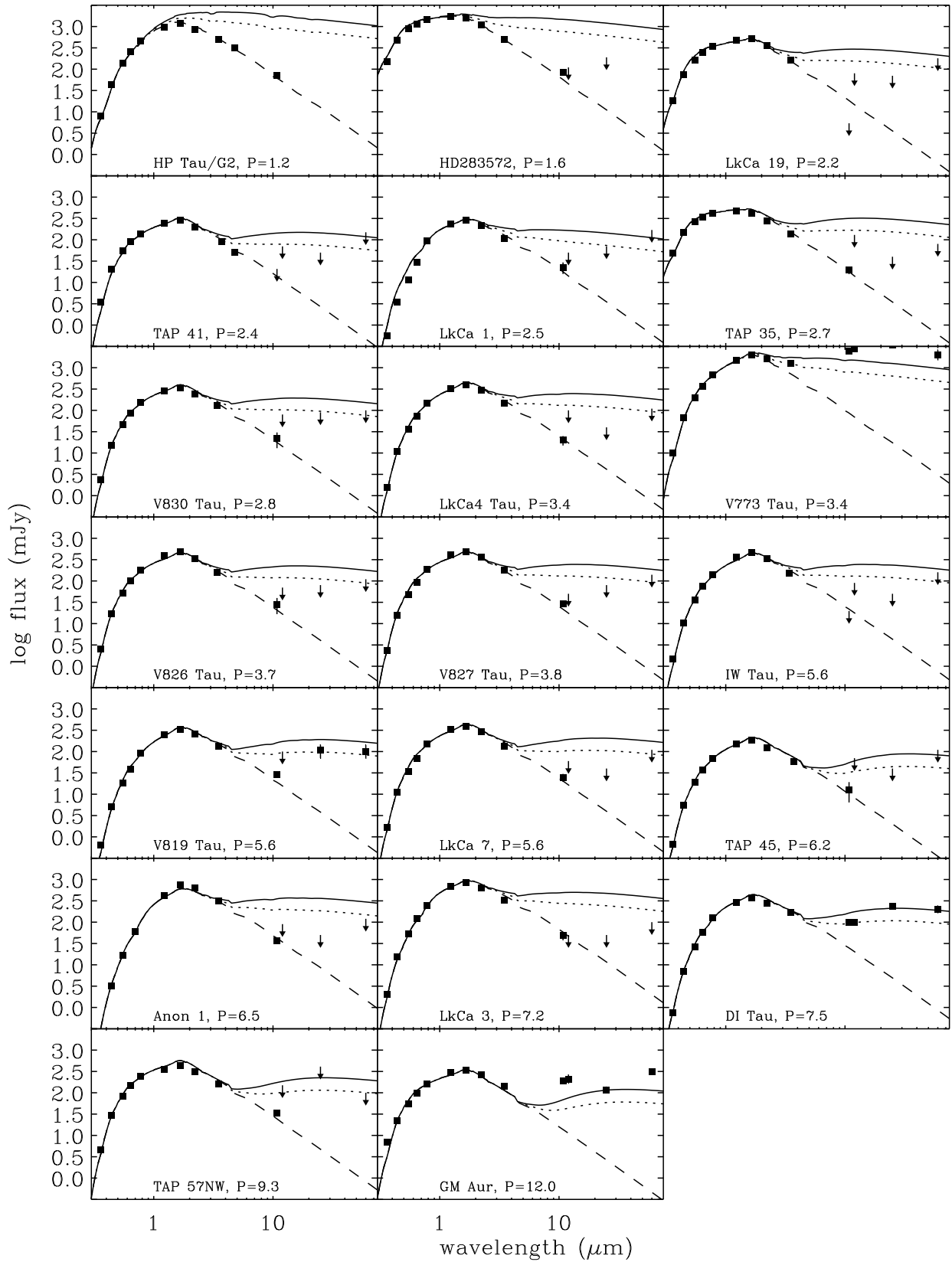
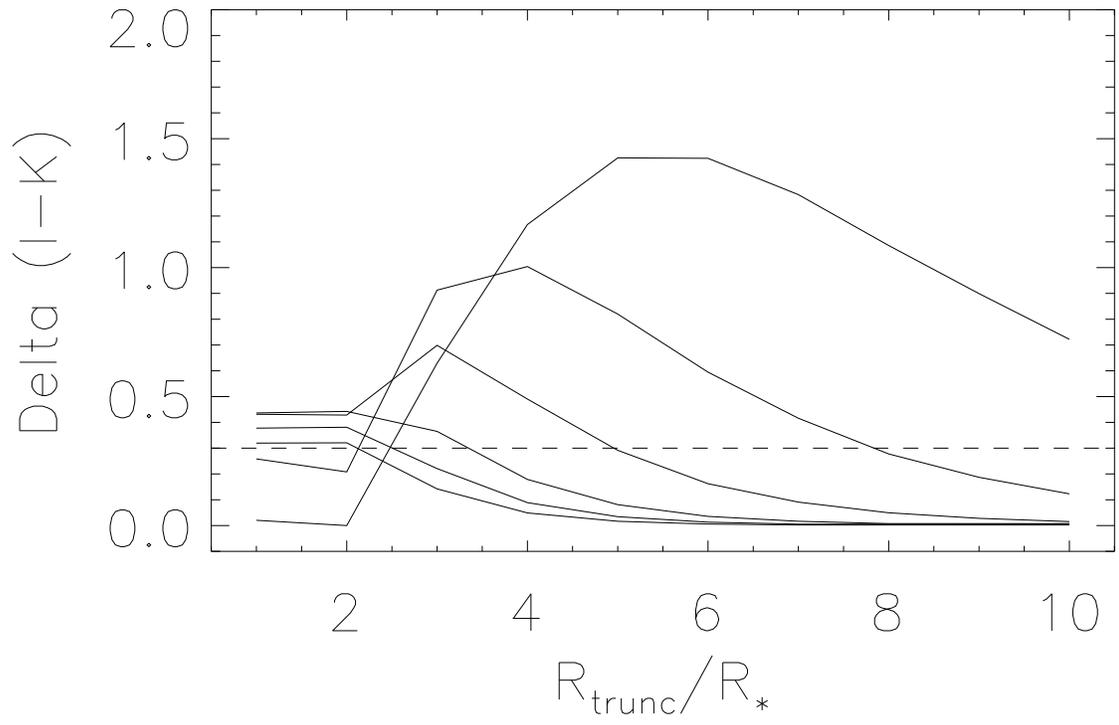
\end{document}